\documentclass[%
 reprint,
 amsmath,assume,
 aps,
]{revtex4-2}

\usepackage{graphicx}
\usepackage{dcolumn}
\usepackage{bm}

\usepackage{color}
\usepackage{pgfplots,mathtools}
\usepackage{url}
\usepackage{hyperref}

\usepackage{braket}
\usepackage{slashed}
\usepackage[compat=1.0.0]{tikz-feynman}
\newcommand{\be}{\begin{equation}}
\newcommand{\ee}{\end{equation}}
\newcommand{\bea}{\begin{eqnarray}}
\newcommand{\eea}{\end{eqnarray}}
\newcommand{\ba}[1]{\begin{array}{#1}}
\newcommand{\ea}{\end{array}}
\newcommand{\nn}{\nonumber}

\newcommand{\del}{\partial}
\newcommand{\orcid}[1]{\href{https://orcid.org/#1}{\includegraphics[width=8pt]{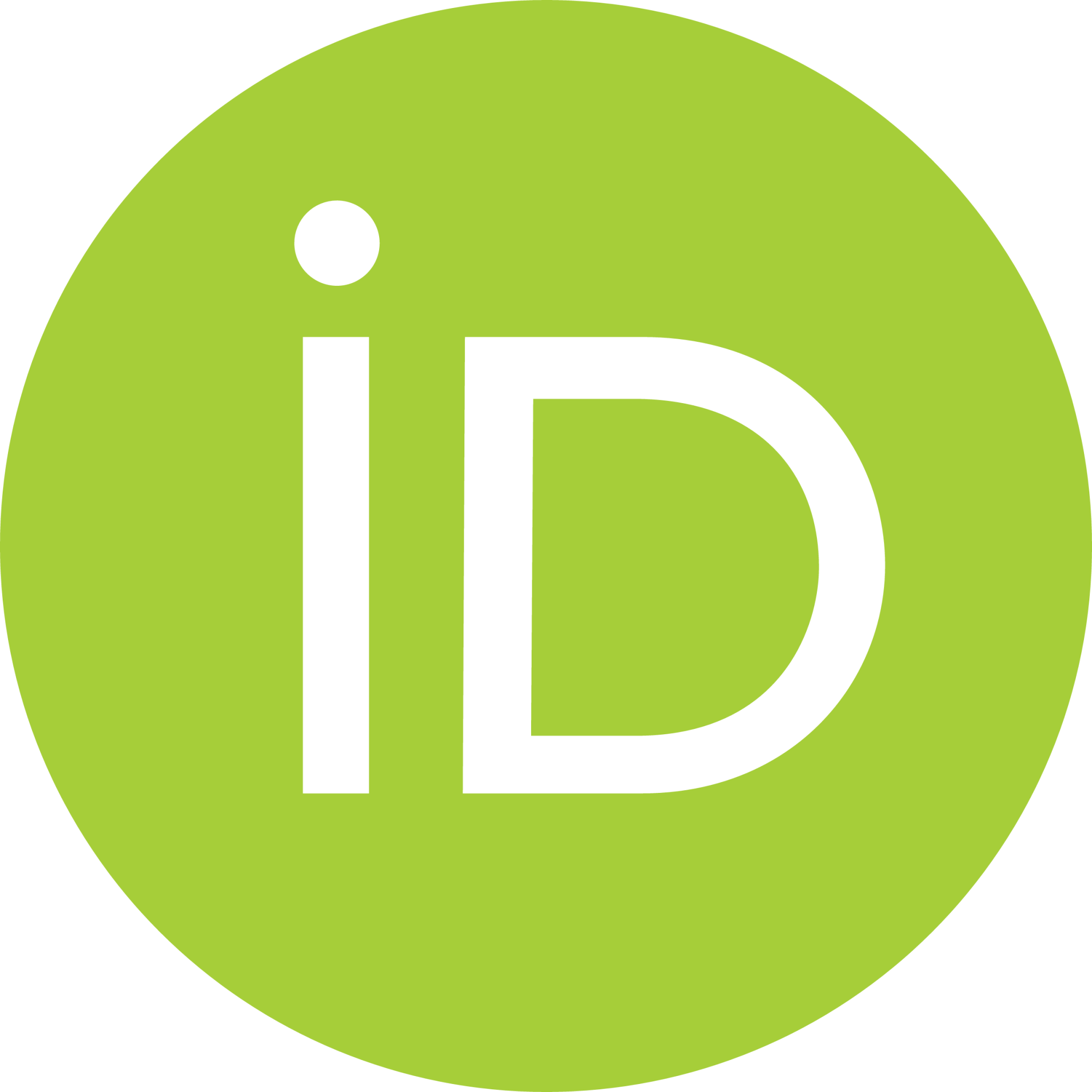}}}

\begin{document}

\preprint{APS/123-QED}

\title{Shear viscosity at finite magnetic field for graphene, non-relativistic and ultra-relativistic cases}
  \author{Cho Win Aung\orcid{0000-0001-5684-2854}~$^{1,2}$, Thandar Zaw Win\orcid{0009-0006-8033-3025}~$^{2}$, Subhalaxmi Nayak\orcid{0009-0001-0145-6785}~$^{2}$, Sabyasachi Ghosh\orcid{0000-0003-1212-824X}~$^{2}$}
  \affiliation{$^{1}$International Institute for Sustainability with Knotted Chiral Meta Matter (WPI-SKCM$^2$),
  	Hiroshima University, 1-3-1 Kagamiyama, Higashi-Hiroshima, Hiroshima 739-8526, Japan}
  
  \affiliation{$^{2}$Department of Physics, Indian Institute of Technology Bhilai,
  	Kutelabhata, Durg 491002, India}
  %
  \begin{abstract}
The present article has addressed the finite magnetic field extension of the previous work by Cho \textit{et al.} (\textit{Phys.
	Rev. B 108}, 235172, 2023) on microscopic calculation of shear viscosity for electron fluid in graphene system. Our calculation is based on the kinetic theory approach in the relaxation time approximation. In the absence of a magnetic field, transport is governed by a single shear viscosity coefficient, whereas the application of a finite magnetic field induces anisotropy, giving rise to five independent shear viscosity coefficients associated with distinct velocity gradient tensors. These coefficients can be physically categorized into perpendicular, parallel, and Hall components relative to the magnetic field direction. When the scattering time equals the cyclotron time, the perpendicular component is suppressed by $80\%$ and the parallel component by $50\%$, and the Hall effect can reach maximum. Corresponding magnetic field strength for electron fluid in graphene is around $0.01-0.1$ Tesla, and the same for non-relativistic electron fluid and ultra-relativistic quark fluid are around $10$ Tesla and $10^{14}$ Tesla respectively. They may be considered as the required magnetic field strength in three different fluid systems to observe noticeable magnetic field response in their shear viscosity coefficients. 
  \end{abstract}
  \maketitle
  \section{Introduction} 
     When there is a conventional flow of electrons in metals, electrons dissipate most of their energy in collisions with the lattice imperfections at low temperatures, called disorder scattering, which are formed because of chemical impurities, vacant lattice sites, and extra atoms not in regular lattice positions. In contrast, the dissipation is mainly in electron-phonon interactions at high temperatures when phonon populations are higher. But, there is a possible electron-electron interaction which is less emphasized in the context of electron transport in metals for a few reasons, primarily due to the relative energy scales, the Pauli exclusion principle, and the screening effects in metals. This kind of condition abruptly changes after the electron-electron scattering becomes prominant since the flow of electrons on 2D graphene layer was found having hydrodynamic behavior \cite{1Ku:2019lgj,2varnavides2020electron,3PhysRevB.103.125106,4PhysRevB.103.235152,5PhysRevB.103.155128,6PhysRevResearch.3.013290,7DiSante:2019zrd,8PhysRevB.103.115402,9sulpizio2019visualizing,10gallagher2019quantum,11doi:10.1126/science.aau0685,12ella2019simultaneous,13bandurin2018fluidity,18doi:10.1126/science.aad0201,14PhysRevB.98.241304,Jaoui2018WP2WF,moll2016evidence,17doi:10.1126/science.aad0343}. See Refs.~\cite{eHD1,eHD2,eHD3, eHD4} for recent review. A macroscopic description of conservation laws in an interacting many-body system like a fluid can be depicted as hydrodynamics. In the hydrodynamics description of Poiseuille flow of electrons, the momentum conserving scattering (electron-electron interaction) dominates rather than the momentum relaxing scattering (electron-phonon, electron-disorder etc.)\cite{PhysRevB.92.115426, Narozhny2017-pa, PhysRevLett.118.226601}. Apart from these recently discovered hydrodynamic properties of electrons in graphene, it was quite famous for its massless nature, concluded from the proportional relation between its energy and momentum\cite{RevModPhys.81.109}. Due to the proportional relation between energy and momentum, electron motion in graphene will not be Galilean-invariant. On the other hand, the relativistic effect of electrons can not be expected, because its velocity ($v\approx 10^6$ m/s) is not very close to the speed of light ($c\approx 3\times 10^8$ m/s). Hence, we can’t  claim the Lorentz-invariant property of electron motion. It opens “unconventional” hydrodynamics\cite{eHD2} as neither non-relativistic hydrodynamics (NRHD) nor relativistic hydrodynamics (RHD) can be applicable. We may call this “unconventional” hydrodynamics as Graphene hydrodynamics (GHD) it is different from the hydrodynamics non-relativistic (NR) and relativistic (R) or ultra-relativistic (UR) cases.
  Now, whenever fluid dynamics or hydrodynamics comes into the picture, then some dissipation coefficient like shear viscosity of that fluid becomes a very important quantity, which does not at all appear in most metals or other condensed matter systems. The present work is aimed at the microscopic calculation of the shear viscosity of this electron fluid in the graphene system, which may be called in short, the graphene fluid (GF). When one microscopically calculates the expression of the shear viscosity of GF, it will be different from its standard expression for non-relativistic fluid (NRF) as well as for relativistic fluid (RF) or ultra-relativistic fluid (URF). Very recently, a relativistic microscopic theory of shear viscosity showing how momentum transport arises from particle dynamics governed by a relativistic Langevin framework has been developed, which smoothly connects with non-relativistic and ultra-relativistic limits~\cite{zaccone2024relativistic}.

  So far, to the best of our knowledge, experimental community~\cite{1Ku:2019lgj,9sulpizio2019visualizing} observed Poiseuille's flow pattern of electrons in graphene, which indirectly reflects the existence of the non-zero viscosity, but the experimental measurement of this shear viscosity coefficient for GF is missing. From the theoretical side, we get only Refs.~\cite{24PhysRevLett.103.025301, Vignale, CWA}, where microscopic expressions of shear viscosity have been addressed. In this context, one can get a long list of Refs.~\cite{QGP1, QGP2, QGP3, QGP4, QGP5, QGP6, QGP7, QGP8, QGP9, QGP10} (and references therein) for microscopical estimations of shear viscosity for relativistic quark and hadronic matter, expected in high-energy heavy ion collision experiments.  Grossly, two classes of frameworks - (1) Kinetic theory approach with relaxation time approximation (RTA)~\cite{QGP1, QGP2, QGP3, QGP4, QGP5, QGP6, QGP7} and (2) Kubo framework~\cite{QGP8, QGP9, QGP10} are adopted by the heavy ion physics community. Both frameworks have a similar structure at the final level expressions for shear viscosity coefficients with two main components. One carries interaction information, called relaxation time, and the remaining part may be called the thermodynamic phase-space of shear viscosity coefficient, which will be a function of temperature and chemical potential. If we analyze the shear viscosity expression of graphene also from Ref.~\cite{24PhysRevLett.103.025301}, then we can identify these two components. The present work has zoomed in on this structure via a systematic calculation of the shear viscosity of GF using the relaxation time approximation method and compared with corresponding structures for NRF and URF. Here, one of our aims is to compare the thermodynamic phase-space component of the shear viscosity coefficient for these three cases - G, NR, and UR.  

  After knowing the lower bound conjecture of shear viscosity to entropy density $(\frac{\eta}{s})$ as $\frac{\hbar}{(4\pi k_B)}$ or $\frac{1}{4 \pi}$ (in natural unit)~\cite{25Kovtun:2004de}, scientific communities are curious to know those strongly coupled systems, which are close to that lower bounds. Experimentally, the RF, like quark and hadronic matter, produced in high-energy heavy ion collision experiments and NRF like cold atom systems~\cite{27Sch_fer_2009}, are identified as those strongly coupled systems. 
  The present article is aimed to calculate shear viscosity coefficient at finite magnetic field for GF and compare with NRF and RF systems. Refs.~\cite{24PhysRevLett.103.025301, CWA} provided a theoretical calculation of shear viscosity for the GF case in the absence of a magnetic field. The present work will explore its finite magnetic field extension. There are rigorous theoretical earlier works \cite{narozhny2019magnetohydrodynamics,alekseev2016negative,PhysRevLett.118.226601} of electron hydrodynamics at finite magnetic field, where they aim to focus on explaining the contemporary experimental phenomenon - negative resistance as a signature of Hall viscosity. In this regard, the present work will not dig into the theoretical framework in that phenomenological direction, rather stay with the comparison type agenda - what is the similarity and dissimilarity in the magnetic field and medium scale of three extreme many-body systems of non-relativistic, relativistic, and graphene cases.

  The article is organized as follows: the first formalism part is addressed in Sec. (\ref{sec: Form}), then the results are given in Sec. (\ref{sec: Results}) and summarized in Sec. (\ref{sec: Sum}). We will use the natural unit $\hbar=c=k_B=1$ in the entire article.
    \section{Formalism}
    \label{sec: Form}
    In the framework of hydrodynamical description, the energy-momentum tensor carries the macroscopic information of any fluid. One can express total energy-momentum tensor $T^{\mu \nu}$ of fluid as a summation of two parts \textendash ~an ideal $T_0^{\mu \nu}$ and a dissipative term $T_D^{\mu \nu}$: 
    \be
        T^{\mu\nu}=T^{\mu\nu}_0 +T^{\mu\nu}_D.
        \label{aa}
    \ee
    The ideal part of the energy-momentum tensor comprises the thermodynamical quantities such as pressure, energy density and number density, etc. In earlier Ref.~\cite{TZW}, we have explored this $T^{\mu\nu}_0$ for graphene hydrodynamics (GHD) by comparing with standard non-relativistic hydrodynamics (NRHD) and relativistic hydrodynamics (RHD). The dissipative part of the energy-momentum tensor $T^{\mu\nu}_D$ carries different dissipative forces or pressure, which are connected with respective transport coefficients like shear viscosity, bulk viscosity, and thermal conductivity. The reader may go through the Ref.~\cite{JD_NJL}
    (and references therein) for this detailed mathematical structure of $T^{\mu\nu}_D$ in the framework of RHD. Among the three transport coefficients, shear viscosity
    coefficient calculation in the GHD framework has recently been obtained in Ref.~\cite{CWA}, where a comparison with NRHD and RHD is also explored.
    Shear viscosity calculations at a finite magnetic field in the RHD framework are done in different Refs.~\cite{Tuchin:2011jw, Ghosh:2018cxb, Mohanty:2018eja, Dey:2019vkn, Dash:2020vxk, Dey:2019axu, Ghosh:2020wqx}. The present work is intended to obtain a similar kind of microscopic calculations of shear viscosity at a finite magnetic field in the GHD framework.

    Before going to discuss  the finite magnetic field framework, let us quickly revisit $T_D^{\mu\nu}$ in the absence of a magnetic field, whose RHD version can be found in Refs~\cite{QGP1, QGP3, QGP8} and GHD version can be found in Ref~\cite{CWA}.
    Considering only shear stress tensor $\pi^{\mu\nu}$ part of $T^{\mu\nu}_D$, we can write
    \bea
    T^{\mu\nu}_D &=& \pi^{\mu\nu} + ...............
    \nn\\
    &=& \eta \mathcal{U}^{\mu\nu} + ...............~,
    \label{Tmn_D}
    \eea
    where  $\eta$ is the shear viscosity coefficient.
    \be 
    \mathcal{U}^{\mu\nu}= D^\mu u^{\nu} +D^\nu u^{\mu} - \frac{2}{3} \Delta^{\mu\nu} \del_\rho u^\rho~,
    \label{u_em1}
    \ee 
    is the fluid velocity gradients, $\Delta^{\mu\nu}=-g^{\mu\nu}+u^\mu u^\nu$ is the projection tensor normal to fluid element velocity $u^\mu$
    and $D^\mu=\del^\mu - u^\mu u^\rho \del_\rho$ is derivative normal to $u^\mu$. Plus dotted in Eq.~(\ref{Tmn_D}) indicate that the dissipation part of energy-momentum tensor has more components associated with bulk viscosity and thermal conductivity, but they
    are not considered here.
    Following the RHD formalism, GHD framework also carry the 4D notation, where greek index like $\mu\equiv(0,i)$ takes values 0 
    for the temporal component and $i=1, 2, 3$ ($i=1, 2$) for the spatial component for 3D (2D) system.
    We can write $\mu$-index based Eqs.~(\ref{Tmn_D}), (\ref{u_em1}) in terms of spatial component $i$-index
    as
    \bea
    T^{ij}_D &=& \pi^{ij}+ .....
    \nn\\
    &=& \eta \mathcal{U}^{ij}+.....~,
    \eea

    with $\mathcal{U}^{ij}=(\del^i u^j +\del^j u^i -\frac{2}{3}\delta^{ij}\nabla \cdot u)$.

    Now, in the presence of a magnetic field, instead of isotropic shear viscosity $\eta$, we will get five different
    components of shear viscosity $\eta_n$, associated with five different (independent) velocity gradients $\mathcal{U}^{ij}_n$ \cite{pitaevskii2017course,Tuchin:2011jw,Ghosh:2018cxb,Mohanty:2018eja,Dey:2019vkn,Dash:2020vxk,Dey:2019axu,HUANG20113075,Ghosh:2020wqx}:
    \bea
        \mathcal{U}_{ij}^0&=&(3b_i b_j-\delta_{ij})(b_k b_lV_{kl}-\frac{1}{3} \vec {\nabla} \cdot \vec{u})\nn\\
        \mathcal{U}_{ij}^1&=&2V_{ij}+\delta_{ij}V_{kl}b_k b_l-2V_{ik}b_j b_k-2V_{jk} b_k b_i\nn\\
        &+&(b_i b_j-\delta_{ij})\vec {\nabla} \cdot\vec{u}\nn\\
        \mathcal{U}_{ij}^2&=&2(V_{ik} b_j b_k+V_{jk} b_i b_k-2V_{kl}b_i b_j b_k b_l)\nn\\
        \mathcal{U}_{ij}^3&=&V_{ik}b_{jk}+V_{jk}b_{ik}-V_{kl}b_{ik}b_jb_l-V_{kl}b_{jk}b_i b_l\nn\\
        \mathcal{U}_{ij}^4&=&2(V_{kl} b_{ik} b_j b_l+V_{kl} b_{jk} b_i b_l)~,
        \label{ff}
    \eea
    where $V_{kl} = \frac{1}{2} (\frac{\partial u_k}{\partial x_l}+\frac{\partial u_l}{\partial x_k})$, $b^{ij}\equiv \epsilon^{ijkl} b_k b_l$, where $\epsilon^{ijkl}$ is an total anti-symmetric Levi-Civita tensor. The reader can see that the velocity gradient components are made by the basic tensors - fluid velocity $(u_i)$, Kronecker delta $(\delta_{i j})$, 
    and the component of magnetic field unit vector,  $b_{i}(B_i\equiv B b_{i})$. 
    For more detailed exploration on this basic physics and mathematics, the reader may go through the Refs.~\cite{pitaevskii2017course,Tuchin:2011jw,Ghosh:2018cxb, Mohanty:2018eja,Dey:2019vkn,Dash:2020vxk,Dey:2019axu, HUANG20113075, Ghosh:2020wqx},
    where the anisotropic nature of shear viscosity components in the presence of magnetic field is well addressed. 
    So, the Newton-Stoke definition for shear viscosity in the presence of magnetic field can be written as
    \be
        \pi_{ij}=\sum_{n=0}^4 \eta_n \mathcal{U}_{ij}^n~.
        \label{pij_Cij}
    \ee

    After introducing the macroscopic Eq.~(\ref{pij_Cij}) of viscous stress tensor $\pi^{ij}$, our next aim will be to find its microscopic expression, and
    at the end we will connect them to obtain the microscopic expression of $\eta_n$. 
    In a microscopic picture, let us consider the electron distribution on a system as a non-equilibrium distribution function $f$, which is a combination of an equilibrium distribution function $f_0$ and the deviation of the equilibrium distribution function $\delta f$ as
    \be
    f=f_0+\delta f~,
    \ee
    where $f_0$ is the same as Fermi-Dirac distribution function $f_0=1/\{\exp{(\beta(E-\mu))}+1\}$. Here, $T=1/\beta$ is the temperature and $\mu$ is the chemical potential or Fermi energy of electrons in the graphene system. Actually, $\mu$ is linked with the net carrier density, which is experimentally tuned via doping method. 
   
    Now, from the kinetic theory, the microscopic expression of the shear stress tensor in graphene can be expressed in terms of $\delta f$ as
    \bea
        &&\pi_{ij}=N_s \int {\frac{d^3 p}{(2\pi)^3}} p_i v_j \delta f~,
        \label{pi_ij}
    \eea
    where $N_s$ is the degeneracy factor or spin weight factor and $v_j=\left(\frac{E}{p^2}\right)p_j$ is the velocity-momentum relation for graphene case. The deviated distribution function can be written as
    \be
    \delta f =\sum_{n=0}^{4}g_n \mathcal{U}_{kl}^n v_k v_l=\sum_{n=0}^{4}g_n \mathcal{U}_{kl}^n \left(\frac{E^2}{p^4}\right)p_k p_l~,
    \label{del_f}
    \ee
    where $g_n$ is the unknown coefficient.
    Substituting the Eq.~(\ref{del_f}) in Eq.~(\ref{pi_ij}) and using the tensor identity, $<p_i p_j p_k p_l>=\frac{1}{15} p^4(\delta_{ij}\delta_{kl}+\delta_{ik}\delta_{jl}+\delta_{il}\delta_{jk})$, the shear stress becomes
    \bea
    &&\pi_{ij}= \frac{N_s}{15}\sum_{n=0}^4 \int {\frac{d^3 p}{(2\pi)^3}} \frac{E^3}{p^2}(2g_n) \mathcal{U}_{ij}^n~.
    \label{pi_ij1}
    \eea
    
    By comparing the microscopic Eq.~(\ref{pi_ij1}) with macroscopic Eq.~(\ref{pij_Cij}), one can get the microscopic expression of shear viscosity coefficient
    \bea
        &&\eta_n=\frac{N_s}{15} \sum_{n=0}^4 \int \frac{d^3 p}{(2\pi)^3} \frac{E^3}{p^2} (2g_n)~.
        \label{eta_n}
    \eea
    To find $g_n$, we use the relaxation time approximation (RTA) based on the Boltzmann transport equation in the presence of the magnetic field
    \cite{pitaevskii2017course,Tuchin:2011jw,Ghosh:2018cxb,Mohanty:2018eja,Dey:2019vkn,Dash:2020vxk,Dey:2019axu,HUANG20113075,Ghosh:2020wqx}:
  \bea
    \frac{\partial f}{\partial t}+\frac{\partial x_i}{\partial t } \frac{\partial f}{\partial x_i}+ \frac{\partial p_i}{\partial t} \frac{\partial f}{\partial p_i}&=&\Big( \frac{\partial f}{\partial t}\Big)_{Col}\\
    \frac{\partial f}{\partial t}+v_i \frac{\partial f}{\partial x_i}+F_i \frac{\partial f}{\partial p_i}&=&\Big( \frac{\partial f}{\partial t}\Big)_{Col}\\
    \frac{\partial f_0}{\partial t}+v_i \frac{\partial f_0}{\partial x_i}+e F_{ij}v_j \frac{\partial \delta f}{\partial p_i}&=& -\frac{\delta f}{\tau_c}~,
    \label{BT}
    \eea
    where $F=e F_{ij}v_j$ is the Lorentz force due to the applied magnetic field, $e$ is the electron's charge, and $\tau_c$ is the relaxation time of the electron. For the finite magnetic field case, we have to consider the electromagnetic field strength tensor $F_{ij}=-Bb_{ij}$, and for the graphene case, we have to use the relation $p_i=\frac{E}{v^2}v_i$. Putting these in the Eq.~(\ref{BT}), we get
    \bea
    &&\frac{E}{v^2 T}v_i v_j f_0(1-f_0) V_{ij}- \frac{e B v^2}{E} b_{ij}v_j\frac{\partial \delta f}{\partial v_i}=-\frac{\delta f}{\tau_c}\\
    &&\frac{E}{v^2 T}v_i v_j f_0(1-f_0) V_{ij}= \frac{1}{\tau_B} b_{ij}v_j\frac{\partial \delta f}{\partial v_i}-\frac{\delta f}{\tau_c}~,
    \label{hh}
    \eea
    where we define the magnetic relaxation or cyclotron time as $\tau_B=\frac{E}{e B v^2}$. By using Eq.~(\ref{del_f}) in Eq.~(\ref{hh}), we get the following equation, 
    \begin{widetext}
    \be
    \frac{E}{v^2 T} v_i v_j V_{ij} f_0(1-f_0) = \frac{2}{\tau_B}\sum_{n=0}^{4}g_n b_{ij}v_j \mathcal{U}_{ik}^n v_k -\frac{1}{\tau_c} \sum_{n=0}^{4}g_n \mathcal{U}_{kl}^n v_k v_l~.
    \label{BTl}
    \ee
    Plugging the Eq.~(\ref{ff}) in the above Eq.~(\ref{BTl}), we get
    \bea
    \frac{E}{v^2 T} v_i v_j f_0(1-f_0) V_{ij}&=&\frac{2}{\tau_B} \Big[g_1 \{2V_{ik} b_{ij} v_j v_k- 2V_{ik} b_{ij}v_j b_k(\vec{v} \cdot \vec{b})\}+g_2 \{2V_{ik} b_{ij} v_j b_k (\vec{v} \cdot \vec{b})\}\nn\\
    &+&g_3 \{2V_{ij} v_i v_j-4V_{ij} v_i b_j (\vec{v} \cdot \vec{b})\}+g_4 \{2V_{ij} v_i b_j (\vec{v} \cdot \vec{b})\} \Big]\nn\\
    &-&\frac{1}{\tau_c} \Big[g_1\{2V_{ij} v_i v_j -4 V_{ij} v_i b_j (\vec{v} \cdot \vec{b})\}+ g_2\{4V_{ij} v_i b_j (\vec{v} \cdot \vec{b})\}\nn\\
    &+&g_3\{2V_{ik} b_{jk} v_i v_j-2V_{ij} b_{ki} v_k b_j (\vec{v} \cdot \vec{b})\}+g_4\{4V_{ij} b_{ki} v_k b_j (\vec{v} \cdot \vec{b})\} \Big]~.
    \eea

    By equating both sides of the above equation, the products of the corresponding tensors can be described as 
    \bea
    v_i v_j V_{ij}&:& \frac{E}{v^2 T} f_0(1-f_0)=-\frac{4g_3}{\tau_B} + \frac{2g_1}{\tau_c}\nn\\
    V_{ij} b_{ik} v_j v_k &:& \frac{4g_1}{\tau_B}+\frac{2g_3}{\tau_c}=0\nn\\
    V_{ij}v_i b_j (\vec{v} \cdot \vec{b})&:& -\frac{8g_3}{\tau_B}+\frac{4g_4}{\tau_B}+\frac{4g_1}{\tau_c}-\frac{4g_2}{\tau_c}=0\nn\\
    V_{ij} b_{ik} v_k b_j (\vec{v} \cdot \vec{b})&:& -\frac{4g_1}{\tau_B}+\frac{4g_2}{\tau_B}-\frac{2g_3}{\tau_c}+\frac{4g_4}{\tau_c}=0~.
    \label{equat_g}
    \eea
    \end{widetext}
    By solving the above four equations, the four coefficients can be obtained as
    \bea
	g_1&=& \frac{E}{2v^2} \beta f_0(1-f_0) \frac{\tau_c}{1+ 4\left(\frac{\tau_c}{\tau_{B}} \right)^2}\nn\\
	g_2&=& \frac{E}{2v^2} \beta f_0(1-f_0) \frac{\tau_c}{1+\left(\frac{\tau_c}{\tau_{B}} \right)^2}\nn\\
	g_3&=& -\frac{E}{v^2} \beta f_0(1-f_0) \frac{\tau_c \left(\frac{\tau_c}{\tau_B} \right)}{1+4\left(\frac{\tau_c}{\tau_{B}} \right)^2}\nn\\
	g_4&=& -\frac{E}{2v^2} \beta f_0(1-f_0) \frac{\tau_c\left(\frac{\tau_c}{\tau_B}\right)}{1+\left(\frac{\tau_c}{\tau_{B}} \right)^2}~.
    \label{const_gs}
    \eea
    To find one of the shear viscosity components, $\eta_0$, one should notice that the equation $\mathcal{U}_{ij}^0 b_i b_j=0$ in which $\mathcal{U}_{ij}^0$ is parallel to the magnetic field direction. Therefore, $\eta_0$ would be unaltered by the magnetic field because of the null contribution of the Lorentz force in the parallel direction. The $\eta_0$ is the shear viscosity in the absence of the magnetic field. The shear viscosity expression with zero magnetic field case is already discussed in our earlier paper \cite{CWA}. It can be rephrased as
    \bea
    \eta_0=\eta &=& \frac{N_s}{15}\int \frac{d^3p}{(2\pi)^3} E^2 \tau_c \beta f_0(1-f_0)
    \nn\\
    &=& \frac{4}{5} \frac{N_s}{\pi^2 v^3} \tau_c T^4 f_4 (A)~,
    \label{eta_0}
    \eea
    where we have used the Fermi-integral identity (see Appendix~ \ref{App_A})
    \be 
    f_{\nu}(A)= \frac{1}{\Gamma_\nu}\int_0^\infty \frac{x^{\nu-1}}{e^x A^{-1}+1} dx~.
    \ee 
    By using Eq.~(\ref{const_gs}) in Eq.~(\ref{eta_n}), we will get the expressions of remaining four shear viscosity coefficients
    \bea
    \eta_1 &=&  \frac{N_s}{15} \int \frac{d^3p}{(2\pi)^3} E^2 \frac{\tau_c}{1+ 4\left(\frac{\tau_c}{\tau_{B}} \right)^2} \beta f_0(1-f_0)~,\nn\\
    \eta_2 &=& \frac{N_s}{15} \int \frac{d^3p}{(2\pi)^3} E^2 \frac{\tau_c}{1+\left(\frac{\tau_c}{\tau_{B}} \right)^2} \beta f_0(1-f_0)~,\nn\\
    \eta_3 &=& -\frac{N_s}{15} \int \frac{d^3p}{(2\pi)^3} E^2 \frac{\tau_c \left(\frac{\tau_c}{\tau_B} \right)}{\frac{1}{2}+2\left(\frac{\tau_c}{\tau_{B}} \right)^2} \beta f_0(1-f_0)~,\nn\\
    \eta_4 &=& -\frac{N_s}{15} \int \frac{d^3p}{(2\pi)^3} E^2 \frac{\tau_c\left(\frac{\tau_c}{\tau_B}\right)}{1+\left(\frac{\tau_c}{\tau_{B}} \right)^2} \beta f_0(1-f_0)~.
    \label{int_four}
    \eea
    By comparing Eq.~(\ref{eta_0}) and Eq.~(\ref{int_four}), one can identify that different components have different effective relaxation times:
    \bea
    \tau_0 &=& \tau_c ~, \nn\\
    \tau_1 &=&  \frac{\tau_c}{1+ 4\left(\frac{\tau_c}{\tau_{B}} \right)^2}~ ,\nn\\
    \tau_2 &=&  \frac{\tau_c}{1+\left(\frac{\tau_c}{\tau_{B}} \right)^2} ~,\nn\\
    \tau_3 &=& \frac{\tau_c \left(\frac{\tau_c}{\tau_B} \right)}{\frac{1}{2}+2\left(\frac{\tau_c}{\tau_{B}} \right)^2} ~,\nn\\
    \tau_4 &=& \frac{\tau_c\left(\frac{\tau_c}{\tau_B}\right)}{1+\left(\frac{\tau_c}{\tau_{B}} \right)^2}~ .
    \label{diff_RT}
    \eea
    For simplicity if we take outside the integration, by considering,
    \begin{equation}
    	\tau_B=\frac{E_{av}}{(v^2eB)}~, \label{tau_B}
    \end{equation}
     with
     \begin{equation}
     	E_{av}=3 T \frac{f_4(A)}{f_3(A)}~, \,    \text{(see Appendix \ref{App_C})} \label{e_av}
     \end{equation} 
      
    we will get,
    \bea
    \eta_1 &=& \frac{4}{5}\frac{N_s}{\pi^2 v^3} \frac{\tau_c}{1+ 4\left(\frac{\tau_c}{\tau_{B}} \right)^2} T^4 f_4 (A)~,\nn\\
    \eta_2 &=& \frac{4}{5}\frac{N_s}{\pi^2 v^3} \frac{\tau_c}{1+\left(\frac{\tau_c}{\tau_{B}} \right)^2} T^4 f_4 (A)~,\nn\\
    \eta_3 &=& -2 \times \frac{4}{5} \frac{N_s}{\pi^2 v^3} \frac{\tau_c \left(\frac{\tau_c}{\tau_B} \right)}{1+4\left(\frac{\tau_c}{\tau_{B}} \right)^2} T^4 f_4 (A)~,\nn\\
    \eta_4 &=& -\frac{4}{5} \frac{N_s}{\pi^2 v^3} \frac{\tau_c\left(\frac{\tau_c}{\tau_B}\right)}{1+\left(\frac{\tau_c}{\tau_{B}} \right)^2} T^4 f_4 (A)~.
    \label{eta_series}
    \eea

    One can use either Eq.~(\ref{int_four}) or  Eq.~(\ref{eta_series}) as both will give similar qualitative dependence of $T$, $\mu$ and $eB$ of different shear viscosity components. For plotting purposes, we will use actual integration results, given in Eq.~(\ref{int_four}), but for interpreting the results, we will use Eq.~(\ref{eta_series}) as reference expressions. The reason is that the Eq.~(\ref{eta_series}) can easily be understood in terms of average $\tau_{B} ~(T, \mu,eB)$ and Fermi integral function $f_4~ (T,\mu)$, and our physics interpretation from the mathematical expressions will be a little less complex.
    \section{Results}
    \label{sec: Results}
    
    Here, in the results section, instead of going to the numerical calculation of the five different components of shear viscosity, let us convert them into three relevant physical components
    - parallel ($\eta_{||}$), perpendicular ($\eta_{\perp}$) and Hall ($\eta_{\times}$) shear viscosity components. Their connecting relations are as follows~\cite{Dey:2019axu, PhysRevD.90.066006, PhysRevD.94.054020}:
      \bea
    \eta_{\perp}     &=& \eta_1 = \frac{1}{1+ 4\left(\frac{\tau_c}{\tau_{B}} \right)^2}\,\eta
    \nn\\
    \eta_{\parallel} &=& \eta_2 = \frac{\tau_c}{1+ \left(\frac{\tau_c}{\tau_{B}} \right)^2}\,\eta
    \nn\\
    \eta_{\times}    &=& \eta_4 = \frac{\frac{\tau_c}{\tau_B}}{1+ \left(\frac{\tau_c}{\tau_{B}} \right)^2}\,\eta~.
    \eea
    Considering the limiting cases for $B\rightarrow 0$ and $\tau_B\rightarrow \infty$ in the Eq.~(\ref{eta_series}), one can easily get 
    the relations $\eta_{\perp}=\eta_{||}=\eta$ and $\eta_{\times}=0$. It means that one can expect an isotropic shear viscosity and zero
    Hall components in the absence of the magnetic field, but when one switches on the magnetic field, non-zero Hall viscosity can be found, as well as 
    an anisotropic nature in the viscosity tensor is also built. 
%
	Adding the hole part with Eq.~(\ref{eta_0}) and~(\ref{eta_series}), we will obtain,
	 
    \bea
    \eta &=&  \frac{4}{5} \frac{N_s}{\pi^2 v^3}\tau_c  T^4 \Big[f_{4}(A)+ f_4 (A^{-1})\Big]~,\\
    \eta_{\perp} &=&  \frac{4}{5}\frac{N_s}{\pi^2 v^3}\frac{\tau_c}{1+4\left(\frac{\tau_c}{\tau_{B}}\right)^2}~ T^4~ \Big[f_{4}(A)+ f_4 (A^{-1})\Big],\\ 
    \eta_{\parallel} &=&  \frac{4}{5}\frac{N_s}{\pi^2 v^3}\frac{\tau_c}{1+\left(\frac{\tau_c}{\tau_{B}} \right)^2} ~T^4~ \Big[f_{4}(A)+ f_4 (A^{-1})\Big]~,\\
    \eta_{\times} &=&  -\frac{4}{5} \frac{N_s}{\pi^2 v^3}\frac{ \tau_c \left(\frac{\tau_c}{\tau_B} \right)}{{1+ \left(\frac{\tau_c}{\tau_{B}} \right)^2}}~ T^4~ \Big[f_{4}(A)+ f_4 (A^{-1})\Big]. 
    \label{eta_series2}
    \eea
    Reader should notice that $\eta_{\perp, \parallel, \times} $ depend on both $\tau_c$ and $\tau_B$ while $\eta$ depends on $ \tau_c$ only.
  Here the magnetic field dependency will be entered through $\tau_{B}$ only and the detailed $T,\mu, eB$ dependent expression for GHD is given in Eq.~(\ref{tau_B}) and~(\ref{e_av}). In place of $E=p v$, if we use ultra- relativistic dispersion relation $E=p$ (by replacing v=1), we will get $\tau_{B}$ for URHD system as,
   \begin{equation}
   	\tau_{B}= \frac{E_{av}}{eB}~,
   \end{equation}
   with
   \begin{equation}
   E_{av}= 3~T ~\frac{f_{4}(A)}{f_{3}(A)}~.
   \end{equation}

   This expression can be used for quark matter, produced in heavy ion collision experiments like LHC or RHIC~\cite{Tuchin:2011jw, Ghosh:2018cxb,Mohanty:2018eja, Dey:2019vkn,Dash:2020vxk, Dey:2019axu, Ghosh:2020wqx, HUANG20113075, pitaevskii2017course, PhysRevD.90.066006, PhysRevD.94.054020}. We will discuss it in more detail gradually. Next, for the NRHD case of electron,

   \begin{equation}
   	\tau_{B}=\frac{m^{*}}{eB}~,
   \end{equation} 
  where $m^{*}$ is an effective electron mass. The effective mass of an electron generally becomes different from its actual mass depending on the many-body effect of different real systems. Depending on real systems, mass can also be constituent particle mass, which may not always be an electron. A detailed calculations of $\eta_n$ for the NRHD case are given in Appendix (\ref{App_B}). Here, we will consider a hypothetical NRHD case with $m^*=m=0.5MeV$(actual electron mass) and interested to compare with realistic GHD case and URHD cases, observed in the Dirac Fluid (DF) domain $\frac{\mu}{T} \ll 1$ of electrons in graphene and free baryon domain of quark matter, produced in LHC and RHIC experiments respectively. Now, if we can analyze $\tau_{B}~(T,\mu, eB)$ for all cases, we can see that all are following inverse proportional dependence with $eB$. We know that quark matter at $\mu=0$ (expected in LHC/ RHIC experiments) follows hydrodynamics very well\cite{27Sch_fer_2009,heinz2013collective,gale2013hydrodynamic}. On the other hand, ultra-clean graphene in the Dirac Fluid domain $\frac{\mu}{T} \ll 1$ or $\mu\rightarrow 0$ shows a prominent electron hydrodynamics\cite{17doi:10.1126/science.aad0343, majumdar2025universality}. So let us keep our concentration at $\mu=0$ for comparing GHD with NRHD and URHD cases. Let us rewrite the cyclotron time or magnetic relaxation time at $\mu=0$,
  \begin{align}
  \tau_{B} &= \frac{m}{eB}  \label{NRtauB}~\text{for NRHD}\\      	
           &= \frac{3~T~ \zeta_4}{\zeta_3~ v^2 ~eB} \label{GtauB} ~  \text{for GHD}\\ 
           &= \frac{3~T~ \zeta_4}{\zeta_3 ~ eB} \label{URtauB}~ \text{for URHD}~. 
  \end{align}
  To estimate $\eta_{\perp, \parallel, \times}$ components for three cases, we have to put some values of $\tau_c$ also along with $\tau_{B}$. In the absence of a magnetic field, one can get the lower limits of $\tau_c$ \cite{CWA} for different cases as:
  \begin{align}
  	\tau_c &= \frac{5}{4 \pi T} \,\text{for GHD/URHD} \label{tauc_GHD_URHD} \\
  	&= \frac{5}{2 \pi T} \,\text{for NRHD} \label{tauc_NRHD}~.
  \end{align}
  From ADS/CFT based calculations \cite{25Kovtun:2004de}, lower bound of $\frac{\eta}{s}$ is $\frac{1}{4 \pi}$, where $s$ is the entropy density. This lower bound is known as the KSS bound or the quantum lower bound. This lower bound conjecture says that $\frac{\eta}{s}$ of any fluid can't be lower than this value, and experimentally observed data follow this conjecture well. This ratio $\frac{\eta}{s}$ of most of the fluid remains quite larger than the KSS bound.
  It is quark matter, produced in HIC experiments, which has $\frac{\eta}{s}$ close to the KSS bound \cite{27Sch_fer_2009}. According to the theoretical calculation of Muller \textit{et al.}~\cite{24PhysRevLett.103.025301}, $\frac{\eta}{s}$ of electron fluid in graphene near charge neutrality point $(\mu \rightarrow 0)$ can reach up to lower value $\frac{5}{4 \pi}$ (five times of KSS bound). In the present article,
   we will theoretically design $\frac{\eta}{s}~(T, \mu=0) = \frac{1}{4 \pi}$ for all cases (NRHD, GHD, and URHD) to explore their extreme fluid scenario, and then we will see their changes due to the magnetic field.
   Our main objective will be to demonstrate the changes or differences in viscosity values due to the magnetic field instead of their absolute values. \\
    
\begin{figure}
	\centering
	\includegraphics[scale= 0.35]{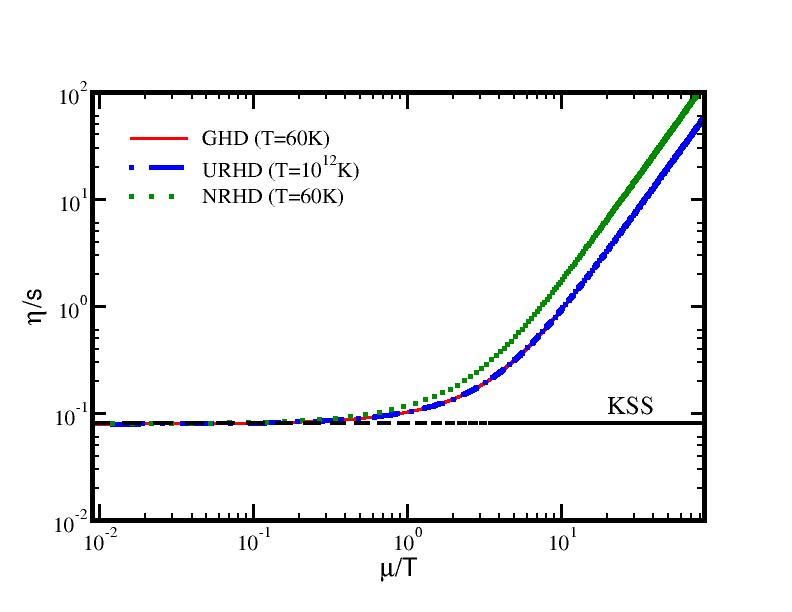}
		\caption{Shear viscosity to entropy density ratio for GHD, URHD and NRHD cases for their corresponding temperatures.}
	\label{fig:1_etabs_muT_3D}
\end{figure}

   Let us plot $\frac{\eta}{s}$ vs $\frac{\mu}{T}$ for NRHD, GHD, and URHD cases in the absence of magnetic field estimation. Fig.(\ref{fig:1_etabs_muT_3D}) demonstrates without magnetic field results of $\frac{\eta}{s}$ vs $\frac{\mu}{T}$, where $\tau_c$'s for different cases are taken from Eq.~(\ref{tauc_GHD_URHD}) and (\ref{tauc_NRHD}), and entropy density expressions~\cite{CWA} are given below,
   
     \begin{align}
   	s_{NR} &=  N_s \Bigg(\frac{m}{2 \pi}\Bigg)^{\frac{3}{2}} T^{\frac{3}{2}} \Bigg[\frac{5}{2}\Bigg(f_{\frac{5}{2}}\left(A\right)  + f_{\frac{5}{2}}\left(A^{-1}\right)\Bigg)\nn \\
   	& -\frac{\mu}{T}\Bigg(f_{\frac{3}{2}}\left(A\right) - f_{\frac{3}{2}}\left(A^{-1}\right)\Bigg)\Bigg]\,\text{for NRHD}~, \label{tau_c_GHD_URHD} \\
   s_G	&= \frac{N_s T^3}{\pi^2 v_g^3}\Bigg[4\Bigg(f_4\left(A\right) + f_4\left(A^{-1}\right)\Bigg)\nn  \\
   & -\frac{\mu}{T}\Bigg(f_3\left(A\right) - f_3\left(A^{-1}\right)\Bigg)\Bigg] \,\text{for GHD}~, \\
  s_{UR}&= \frac{N_s T^3}{\pi^2}\Bigg[4\Bigg(f_4\left(A\right) + f_4\left(A^{-1}\right)\Bigg)\nn \\
   & -\frac{\mu}{T}\Bigg(f_3\left(A\right) - f_3\left(A^{-1}\right)\Bigg)\Bigg] \,\text{for URHD}~. \label{s_}
   \end{align}
Expression of $\eta$ for URHD will be the same as Eq.~(\ref{eta_0}) just by replacing $v=1$ and $\eta$ for NRHD is given in Eq.~(\ref{eta_0_nr}) of the Appendix, where electron mass $m=0.5~MeV$ is taken. Temperature $T$ is kept fixed at $T=60K \approx 5~meV$ (milli-electron Volt) for NRHD and GHD, while for URHD, it is kept at $T= 10^{12}K \approx 300~MeV$ (mega-electron Volt) (temperature of quark matter). So, almost two extreme $T-\mu$ domains- (1) Condensed Matter Physics (CMP) domain (NRHD and GHD) and (2) High Energy Physics (HEP) domain (URHD) are our matter of interest, which is nicely pictorized in Fig.(1) of Ref.~\cite{CWA}.\\
For the CMP domain, $\mu$ denotes the Fermi energy of electrons which will vary in eV units with the help of the doping technique in the experiment. For GHD, Dirac Fluid(DF) and Fermi Liquid(FL) domains are defined by $\frac{\mu}{T} \ll 1$ and $\frac{\mu}{T} \gg 1$ regions. Reader may remember that in metal, electron Fermi energy range is $\mu=2-10 ~eV$, for which $\frac{\mu}{T} \approx 400-2000~ (T=60K)$ or $80-400$~(even at room temperature $T=300K$), i.e. $\frac{\mu}{T} \gg 1$, and so FL or Fermi gas (FG) domain is well satisfied. That is why the electron Fermi gas is well applicable for the metal case.\\

Now, let us come to the HEP domain, where $\mu$ is the quark chemical potential and will vary in MeV units. Interestingly, the dimensionless quantity $\frac{\eta}{s}$ for CMP and HEP domains shares an approximate similarity in qualitative $\frac{\mu}{T}$ dependence and order of magnitude, although their absolute $\mu-T$ values are quite different. This beauty is coming because of the thermodynamical phase space, governed by Fermi-Dirac statistics. The growing trend of $\frac{\eta}{s}$ beyond $\frac{\mu}{T}=1$ reflects the transition from fluid to non-fluid nature, which is well observed in the graphene system. Lower values of $\frac{\eta}{s}$  correspond to higher interaction cross-section, which will develop a strongly coupled/interacting many-body system. On the other hand, higher values of $\frac{\eta}{s}$ are associated with  a weakly couple/interacting gas picture. If the weakness in interaction grows too high to create a velocity gradient of fluid, then the fluid will behave like non-fluid many-body system. In graphene, we get this fluid to non-fluid transition signature when one transit from DF ($\frac{\mu}{T} \ll 1$) to FL ($\frac{\mu}{T} \gg 1$) domain by tuning $\mu$ and $T$ via doping methods~\cite{17doi:10.1126/science.aad0343, majumdar2025universality}. Quark fluid may face similar transition like DF to FL as we transist from small quark chemical potential ($\frac{\mu}{T} \ll 1$) to its large values ($\frac{\mu}{T} \gg 1$)\cite{rai2025towards}. That is an interesting matter of research in the HEP community if they take GHD observation in the CMP domain as a guiding reference. The present article is not intended to explore this fluid to non-fluid transition aspect, so we will stick to the DF domain ($\mu = 0, \mu \neq 0$ but $\frac{\mu}{T} < 1$), where shear viscosity will not be undefined.
 \begin{figure}
	\centering
	\includegraphics[scale= 0.35]{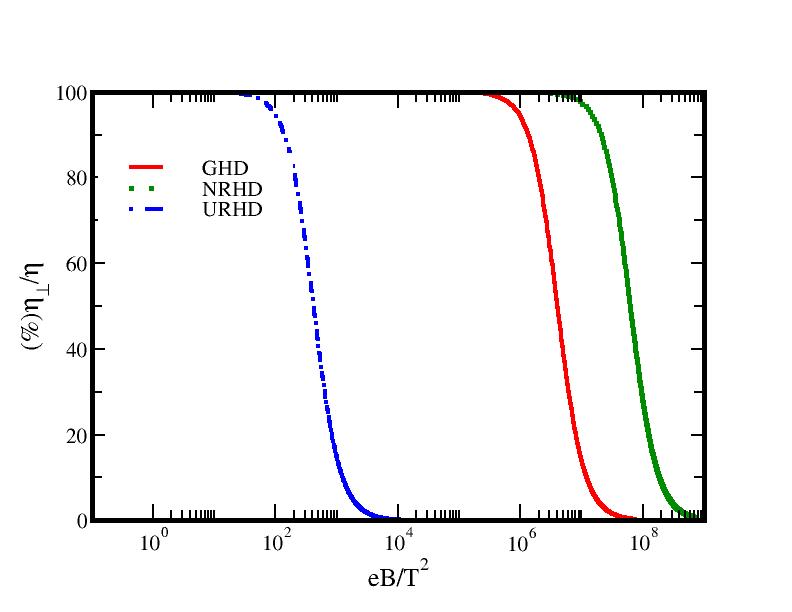}
		\caption{Magnetic field dependent of perpendicular shear viscosity ratio in graphene, non-relativistic and ultra-relativistic systems.}
	\label{fig:eta_perpen}
\end{figure}
\begin{figure}
	\centering
	\includegraphics[scale= 0.35]{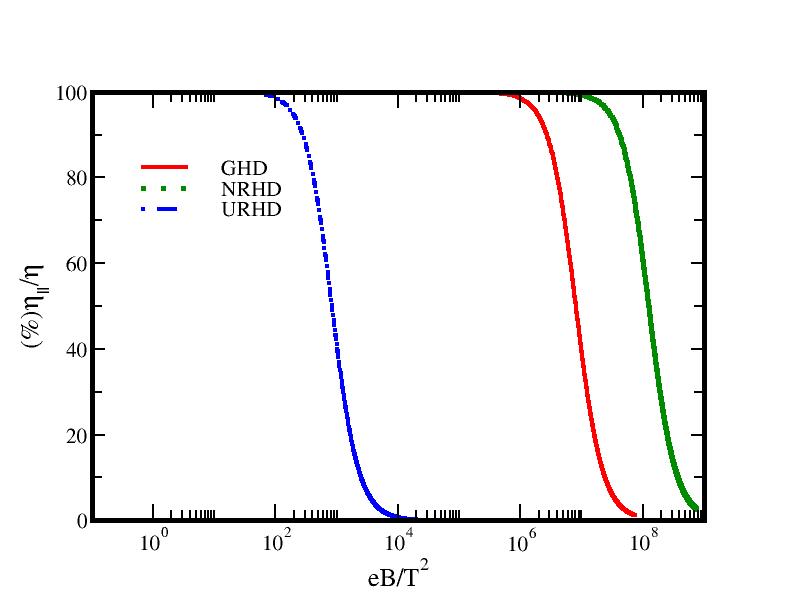}
		\caption{Magnetic field dependent of parallel shear viscosity ratio in graphene, non-relativistic and ultra-relativistic systems.}
	\label{fig:eta_para}
\end{figure}

\begin{figure}
	\centering
	\includegraphics[scale= 0.35]{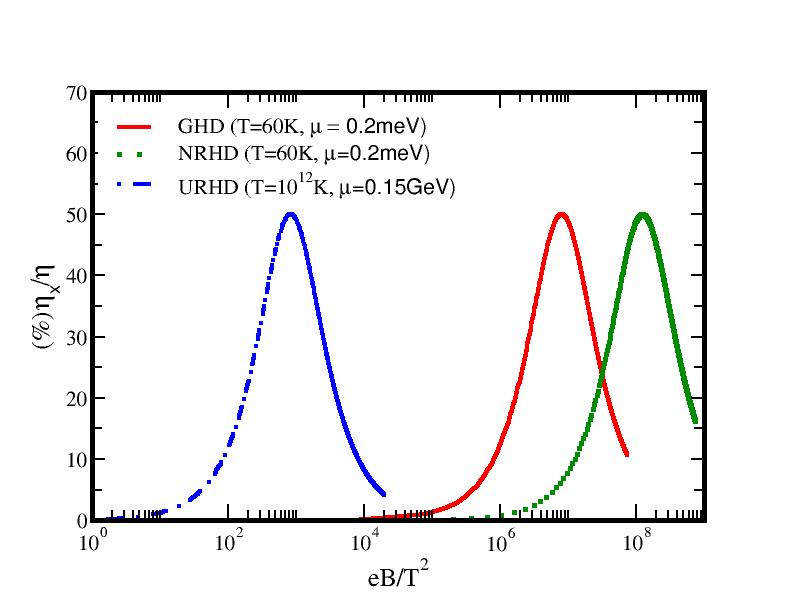}
		\caption{Magnetic field dependent of Hall viscosity ratio in graphene, non-relativistic and ultra-relativistic hydrodynamic regimes. GHD and NRHD results are shown as at $T=60K$ and $\mu=0.2~meV$ while URHD case corresponds to $10^{12}K$ and $\mu=0.15~GeV$.}
	\label{fig:eta_hall}
\end{figure}
Next, Fig.(\ref{fig:eta_perpen}), (\ref{fig:eta_para}) and (\ref{fig:eta_hall}) demonstrate the $\frac{eB}{T^2}$ dependence of normalized shear viscosity components $\frac{\eta_{\perp}}{\eta}$, $\frac{\eta_{\parallel}}{\eta}$ and $\frac{\eta_{\times}}{\eta}$  at $\mu=0.002~eV$  (in milli-electron Volt range), from where a percentage changes can be noticed. We have chosen $\mu=0.002~eV$ to zoom in on the non-zero Hall component, which becomes zero at $\mu=0$ due to the cancellation of electron and hole contributions. Through a simple mathematical analysis, when $\tau_c=\tau_B$, the parallel and perpendicular components will get $50\%$ and $80\%$ reduction, and the Hall component becomes maximum. From Fig.(\ref{fig:eta_perpen})  and (\ref{fig:eta_para}), we notice that for NRHD case, we don't see any impact of magnetic field within the range $\frac{eB}{T^2}=0-10^7$ as we get $\eta_{\parallel} \approx \eta_{\perp}$ $\approx$ $\eta$. Beyond this range $(eB>10^7 ~T^2)$, we can get the anisotropy $(\eta_{\parallel} \neq \eta_{\perp})$ of shear viscosity tensor and with magnetic field strength viscosity components are reduced by following a ranking $\eta_{\perp} < \eta_{\parallel} < \eta$. So, there is a threshold value of $eB$ in terms of $T^2$  units, beyond which the impact of the magnetic field can be noticed, and an anisotropy property in the shear viscosity tensor will be built. These threshold values of GHD and URHD are different ($\frac{eB}{T^2} \approx 10^5$ and $10$ respectively). The Hall viscosity component will be maximum near these threshold values for different cases. Using Eqs.~(\ref{NRtauB}), (\ref{GtauB}), (\ref{URtauB}) and  Eqs.~( \ref{tauc_GHD_URHD}) (\ref{tauc_NRHD}), we can get the order of magnetic field strength, required for maximum Hall viscosity.\\

For NRHD case,
 \begin{align}
	\frac{\tau_{c}}{\tau_{B}} &= 1 \nn \\      	
  \implies \frac{5}{2 \pi T} \times \frac{eB}{m}	&= 1 \nn \\ 
\implies \frac{eB}{T^2} 	&= \frac{2 \pi}{5} \Big(\frac{m}{T}\Big) , \nn \\
 &\approx 1.2 \times 10^8 \, \nn \\
  & \text{(using  T= 5meV and 
  m=0.5MeV)}  \nn \\
 \implies eB &\approx 10\,  Tesla~.
 \label{NRHD_eB}
\end{align}

For GHD case,
\begin{align}
	\frac{\tau_{c}}{\tau_{B}} &= 1 \nn \\      	
	\implies \frac{5}{4 \pi T} \times \frac{ v^2 eB}{3T \frac{\zeta_4}{\zeta_3}}	&= 1 \nn \\ 
	\implies \frac{eB}{T^2} 	&= \frac{12 \pi}{5} \, \frac{\zeta_4}{\zeta_3} \frac{1}{v^2} \nn \\
	&\approx 10^5 - 10^6 \, \text{(using v=0.003 - 0.01)} \nn \\
	\implies eB &\approx 10^{-2} - 10^{-1}\,  Tesla~.
	\label{GHD_eB}
\end{align}

For URHD case,
\begin{align}
	\frac{\tau_{c}}{\tau_{B}} &= 1 \nn \\      	
	\implies \frac{5}{4 \pi T} \times \frac{eB}{3T \frac{\zeta_4}{\zeta_3}}	&= 1 \nn \\ 
	\implies \frac{eB}{T^2} 	&= \frac{12 \pi}{5} \, \frac{\zeta_4}{\zeta_3}  \nn \\
	&\approx 6.8  \, \text{(using T=0.1 - 0.4 GeV)} \nn \\
	\implies eB &= 4.3m_\pi^2 - 68.6m_\pi^2 \nn \\
	&\approx 10^{14}-10^{15} ~Tesla.
	\label{URHD_eB}
\end{align}
Above approximation calculations in Eqs.~\eqref{NRHD_eB} \eqref{GHD_eB}, and \eqref{URHD_eB} demonstrate $\frac{eB}{T^2}\approx10^8$ (NRHD), $10^6$ (GHD), and $10$ (URHD) are the positions for getting maximum Hall viscosity, which are little shifted in actual plot in Fig.~(\ref{fig:eta_hall}). There are two sources of reason for this deviation. One is for the difference between interpreting Eq.~(\ref{eta_series}) and result-generating Eq.~(\ref{int_four}). Another is the difference between $\mu \rightarrow 0$ approximation of Eqs.~\eqref{NRHD_eB}, \eqref{GHD_eB}, \eqref{URHD_eB} and $\mu \neq 0$ actual estimation from Eq.~(\ref{int_four}), where hole contributions are added to generate the final graph in Fig.~(\ref{fig:eta_hall}).\\

Results of the GHD system show that the maximum Hall viscosity strength or $80\%$ reduction of perpendicular viscosity component can be achieved only using a very accessible field strength $10^{-2}-10^{-1}$ Tesla, which can be experimentally realized by measuring the geometrical changes in the parabolic pattern of velocity profile.
Results of the URHD system indicate that the quark matter, produced in peripheral HIC experiments, where $eB=1-20 m_{\pi}^2$~\cite{Tuchin:2011jw} is expected to be produced, can reach the highest Hall viscosity and $80\%$  reduction of perpendicular viscosity.\\
Our results can be applicable for more detailed phenomenological studies of QGP viscosity at finite magnetic field as done by earlier Refs.~\cite{Tuchin:2011jw,Ghosh:2018cxb,Mohanty:2018eja,Dey:2019vkn,Dash:2020vxk, Dey:2019axu,Ghosh:2020wqx,HUANG20113075,pitaevskii2017course,PhysRevD.90.066006,PhysRevD.94.054020}. For the GHD case also, one can relate it with experimental results obtained by Ref.~\cite{11doi:10.1126/science.aau0685} after transforming the calculation to 2D geometry.

    \section{Summary and Interpretation}
    \label{sec: Sum}
 In summary, we have calculated the shear viscosity tensor of the electron fluid in the graphene system at a finite magnetic field using the relaxation time approximation based on the kinetic theory framework. In contrast to the isotropic shear viscosity coefficient in the absence of a magnetic field, an anisotropic transportation will be created due to a magnetic field, which gives rise to five independent shear viscosity coefficients accompanied by five velocity gradient tensors. These coefficients are physically categorized by three components: perpendicular, parallel, and Hall components according to the applied magnetic field direction. At the zero magnetic field limit, there will be no Hall viscosity and no anisotropy, which means that the parallel and perpendicular components remain the same and equal to the isotropic value. As the magnetic field increases, anisotropy in the viscous tensor starts, which means that parallel and perpendicular components of viscosity become unequal. However, for building a noticeable anisotropy, the magnetic field should cross a threshold value, which is around $10^{-2}-10^{-1}$ Tesla, $10$ Tesla, and $10^{14}-10^{15}$ Tesla for electron fluid in graphene system, non-relativistic electron fluid and ultra-relativistic quark/hadronic fluid respectively. Near the charge neutrality point $(\frac{\mu}{T} \rightarrow 0)$, the Hall viscosity of the electron fluid in graphene disappears due to the cancellation of the electron and hole contributions. The same is true for quark/hadronic fluid at zero (net) quark/baryon density or chemical potential, expected in RHIC and LHC experiments. So, for getting a non-zero Hall viscosity, we need a finite electron Fermi energy and quark chemical potential for which electron and quark density become larger than the hole and anti-quark density, respectively. The former case for the graphene system can be achieved via the doping method, and the latter case for quark/hadronic fluid at finite quark/baryon chemical potential can be produced via low center of mass energy collision experiments like CBM and NICA. However, there is a possibility of breaking down the fluid property due to the fluid to non-fluid transition as the density increases. So, to see non-zero Hall viscosity, we have to stay in such a small but non-zero density, where the fluid property does not break down. When the relaxation time scale is equal to the cyclotron time scale, the perpendicular component is reduced to $80\%$, the parallel component is reduced to $50\%$, and the Hall is showing a maximum. To see the noticeable anisotropy and maximum Hall viscosity, different fluid systems need different magnetic field strengths as mentioned earlier, where interesting news is that graphene requires only $0.01-0.1$ Tesla to observe maximum Hall viscosity.     
    \begin{acknowledgments}
    This work was partly  supported by the WPI program “Sustainability with Knotted Chiral Meta Matter (SKCM2)” at Hiroshima University, Japan (C.W.A), the Doctoral Fellowship in India (DIA) program of the Ministry of Education (MoE), Govt. of India (T.Z.W.), the Ministry of
    Education (MoE), Govt. of India (S.N.) and the Board of Research in Nuclear Sciences (BRNS)
    and Department of Atomic Energy (DAE), Govt. of India, under Grant No. 57/14/01/2024-BRNS/313 (S.G.). The authors thank Dr. Aritra Bandyopadhyay (West University of Timisoara, Romania) for fruitful discussions. We also thank the other members of eHD club (Sesha P. Vempati, Ashutosh Dwibedi, and Narayan Prasad).
    \end{acknowledgments}
    \appendix
    \section{Fermi integral for 3D graphene}
    \label{App_A}
    For 3D graphene shear viscosity components calculation, the common integration part will be 
    \bea
    \int_0^\infty d^3p~ E^2 \beta f_0(1&-&f_0) = \frac{4 \pi}{v^3} \int_0^\infty E^4 dE  \frac{\partial f_0}{\partial \mu}\nn\\
    &=& \frac{4 \pi}{v^3} \frac{\partial}{\partial \mu} \int_0^\infty \frac{E^4}{e^{\beta(E-\mu)}+1} dE \nn\\
    &=& \frac{4 \pi}{v^3} \frac{1}{\beta^5} \frac{\partial}{\partial \mu} \int_0^\infty \frac{x^{5-1}}{e^x A^{-1}+1} dx.
    \label{com}
    \eea
    where we use $x=\beta E$, $A=\exp{(\beta \mu)}$ and the identity, $\beta f_0(1-f_0)=\frac{\partial f_0}{\partial \mu}$. By utilizing the Fermi-Dirac integral function, $f_{\nu}(A)= \frac{1}{\Gamma_\nu}\int_0^\infty \frac{x^{\nu-1}}{e^x A^{-1}+1} dx$ with its derivative, $\frac{\partial }{\partial \mu}f_{\nu}(A)= \beta f_{\nu-1}(A)$, one can rewrite the Eq.~(\ref{com}) as
    \be
    \int_0^\infty d^3p~ E^2 \beta f_0(1-f_0) = \frac{4 \pi}{v^3} \times 24 ~T^4 f_4(A)~.
    \label{int_fun}
    \ee
    By inserting the Eq.~(\ref{int_fun}) in Eq.~(\ref{int_four}), we get the shear viscosity coefficients in terms of Fermi functions as in Eq.~(\ref{eta_series}).
    
    \section{Shear viscosity expression for 3D non-relativistic system}
    \label{App_B} 
    Since the mass-momentum relation is $p_i=m v_i$ for the non-relativistic system, Eq.~(\ref{del_f}) becomes 
    \bea
        \delta f&=&\sum_{n=0}^{4}g_n \mathcal{U}_{kl}^n v_k v_l=\sum_{n=0}^{4}g_n \mathcal{U}_{kl}^n \frac{p_k p_l}{m^2}~.
        \label{app2}
    \eea
    The non-relativistic description of Eq.~(\ref{pi_ij1}) can be described as
    \bea
        &&\pi_{ij}= \frac{2 N_s}{15 m^3} \sum_{n=0}^4 \int {\frac{d^3 p}{(2\pi)^3}} p^4 g_n \mathcal{U}_{kl}^n~.
        \label{app4}
    \eea
    The Eq.~(\ref{eta_n}) comes out as
    \bea
        &&\eta_{n}= \frac{2N_s}{15 m^3} \sum_{n=0}^4 \int {\frac{d^3 p}{(2\pi)^3}} p^4 g_n~.
        \label{app4}
    \eea
    In order to find $g_n$, the BTE for the non-relativistic system is
    \begin{widetext}
    \bea
    \frac{m}{T} v_i v_j  f_0(1-f_0) V_{ij} &=& \frac{2}{\tau_B} \sum_{n=0}^4 g_n b_{ij}v_j \mathcal{U}_{ik}^n v_k -\frac{1}{\tau_c} \sum_{n=0}^4 g_n \mathcal{U}_{kl}^n v_k v_l
    \label{app6}
    \eea
    where the cyclotron time is $\tau_B=\frac{m}{e B}$ for non-relativistic system.
    \bea
    \frac{m}{T} v_i v_j V_{ij} f_0(1-f_0)&=&\frac{2}{\tau_B} \Big[g_1 \{2V_{ik} b_{ij} v_j v_k- 2V_{ik} b_{ij}v_j b_k(\vec{v} \cdot \vec{b})\}+g_2 \{2V_{ik} b_{ij} v_j b_k (\vec{v} \cdot \vec{b})\}\nn\\
    &+&g_3 \{2V_{ij} v_i v_j-4V_{ij} v_i b_j (\vec{v} \cdot \vec{b})\}+g_4 \{2V_{ij} v_i b_j (\vec{v} \cdot \vec{b})\} \Big]\nn\\
    &-&\frac{1}{\tau_c} \Big[g_1\{2V_{ij} v_i v_j -4 V_{ij} v_i b_j (\vec{v} \cdot \vec{b})\}+ g_2\{4V_{ij} v_i b_j (\vec{v} \cdot \vec{b})\}\nn\\
    &+&g_3\{2V_{ik} b_{jk} v_i v_j-2V_{ij} b_{ki} v_k b_j (\vec{v} \cdot \vec{b})\}+g_4\{4V_{ij} b_{ki} v_k b_j (\vec{v} \cdot \vec{b})\} \Big]~.
    \label{app7}
    \eea   
    \end{widetext}
    Now, the Eq.~(\ref{equat_g}) becomes as
    \bea
    v_i v_j V_{ij} &:& \frac{m}{T} f_0(1-f_0)=\frac{4g_3}{\tau_B} - \frac{2g_1}{\tau_c}\nn\\
    V_{ij} b_{ik} v_j v_k &:& \frac{4g_1}{\tau_B}+\frac{2g_3}{\tau_c}=0\nn\\
    V_{ij}v_i b_j (\vec{v} \cdot \vec{b})&:& -\frac{8g_3}{\tau_B}+\frac{4g_4}{\tau_B}+\frac{4g_1}{\tau_c}-\frac{4g_2}{\tau_c}=0\nn\\
    V_{ij} b_{ik} v_k b_j (\vec{v} \cdot \vec{b})&:& -\frac{4g_1}{\tau_B}+\frac{4g_2}{\tau_B}-\frac{2g_3}{\tau_c}+\frac{4g_4}{\tau_c}=0.
    \label{app8}
    \eea
    The Eq.~(\ref{const_gs}) is given by
    \bea
	g_1&=& \frac{m}{2T} f_0(1-f_0) \frac{\tau_c}{1+ 4\left(\frac{\tau_c}{\tau_{B}} \right)^2}~,\nn\\
	g_2&=& \frac{m}{2T} f_0(1-f_0) \frac{\tau_c}{1+\left(\frac{\tau_c}{\tau_{B}} \right)^2}~,\nn\\
	g_3&=& -2 \ \frac{m}{2T} f_0(1-f_0) \frac{\tau_c \left(\frac{\tau_c}{\tau_B} \right)}{1+4\left(\frac{\tau_c}{\tau_{B}} \right)^2}~,\nn\\
	g_4&=& -\frac{m}{2T} f_0(1-f_0) \frac{\tau_c\left(\frac{\tau_c}{\tau_B}\right)}{1+\left(\frac{\tau_c}{\tau_{B}} \right)^2}~.
    \label{app9}
    \eea
    The shear viscosity coefficient in the absence of magnetic field is
    \be
    \eta_{0}=\eta= N_s \left( \frac{m}{2 \pi}\right)^{3/2} \tau_c T^{5/2} f_{5/2}(A)~.
    \label{eta_0_nr}
    \ee
    The non-relativistic version of Eq.~(\ref{eta_series}) in terms of the Fermi function can be expressed as
    \bea
    &&\eta_{1}= N_s \left( \frac{m}{2 \pi}\right)^{3/2} \frac{\tau_c}{1+ 4\left(\frac{\tau_c}{\tau_{B}} \right)^2} T^{5/2} f_{5/2}(A)~,\nn\\
    &&\eta_{2}= N_s \left( \frac{m}{2 \pi}\right)^{3/2} \frac{\tau_c}{1+ \left(\frac{\tau_c}{\tau_{B}} \right)^2} T^{5/2} f_{5/2}(A)~,\nn\\
    &&\eta_{3}= -2 N_s \left( \frac{m}{2 \pi}\right)^{3/2} \frac{\tau_c \left(\frac{\tau_c}{\tau_B} \right)}{1+ 4\left(\frac{\tau_c}{\tau_{B}} \right)^2} T^{5/2} f_{5/2}(A)~,\nn\\
    &&\eta_{4}= -N_s \left( \frac{m}{2 \pi}\right)^{3/2} \frac{\tau_c \left(\frac{\tau_c}{\tau_B} \right)}{1+ \left(\frac{\tau_c}{\tau_{B}} \right)^2} T^{5/2} f_{5/2}(A)~.
    \label{app11}  
    \eea
    \section{Average energy expression for 3D graphene system}
    \label{App_C} 
    If the total number of energy states in energy range $E$ to $E + dE$ are $D\left(E\right)dE$, then the density of states will be 
    \begin{equation}
    g\left(E\right) = \frac{D\left(E\right)dE}{dE}~.
    \end{equation}
    For the 3D graphene system, the number of energy states is
    \begin{equation}
    D\left(E\right)dE = N_s \frac{4\pi V}{h^3 v^3}E^2 dE~.
    \label{eng_stat}
    \end{equation}
    The total number of electrons at any value of temperature is given by,
    \begin{equation}
    N = \int_0^{\infty}D\left(E\right)dE f_0~.
    \label{3dnd}
    \end{equation}
    After plugging the value of Eq.~(\ref{eng_stat}) in the above Eq.~(\ref{3dnd}) and $f_0$, the total number of electrons is,
    \begin{align*}
    &N = N_s \frac{4\pi V}{\left(2\pi\right)^3 v^3} \int_{0}^{\infty} \frac{E^2}{e^{\beta E} A^{-1}+ 1} dE~,
    \end{align*}
    where $V$ is the volume of the system. The expression of number density as 
    \begin{equation}
    n = \frac{N}{V} = \frac{N_s}{\pi^2 v^3} T^3 f_3(A)~.
    \end{equation}
    Similarly, the total energy of electrons is given by,
    \begin{equation}
    U = \int_0^{\infty}D\left(E\right) [E] dE f_0~.
    \label{ener_gra}
    \end{equation}
    Now from the Eq.~(\ref{3dnd}), the energy of electrons is,
    \begin{equation}
    U= N_s \frac{4\pi V}{\left(2\pi\right)^3 v^3} \int_0^{\infty} \frac{E^3}{e^{\beta E} A^{-1}+ 1} dE~.
    \end{equation}
    The energy density becomes as
    \begin{equation}
    \epsilon = T_0^{0 0} =  N_s \frac{3}{\pi^2 v^3} T^4 f_4(A).
    \end{equation}
    The average energy of 3D graphene is 
    \bea
    E_{av}&=& \frac{U}{N}=\frac{\epsilon}{n}=3 T \frac{f_4(A)}{f_3(A)}~,
    \label{avgE_G}
    \eea
    where the Fermi-Dirac integral is given by
    \begin{equation}
    f_\nu (A)=\frac{1}{\Gamma (\nu)}\int_0^\infty \frac{x^{\nu-1}}{A^{-1} e^x+1} dx~.
    \label{mom}
    \end{equation}
    
    \bibliographystyle{unsrt}
    \bibliography{ref}
    \end{document}